\documentstyle[twoside,fleqn,espcrc2,epsf]{article}

\title{Center vortex model for the infrared sector of Yang-Mills 
theory\thanks{Talk presented by M. Engelhardt} }

\author{M. Engelhardt\address[TUB]{Institut f\"ur Theoretische Physik,
Universit\"at T\"ubingen, \\
Auf der Morgenstelle 14, 72076 T\"ubingen, Germany}\thanks{Supported
by DFG under En 415/1-1, Re 856/4-1 and Schm 1342/3-1.},
M. Faber\address{Atominstitut der \"osterreichischen Universit\"aten,
Arbeitsgruppe Kernphysik, \\
Technische Universit\"at Wien, A-1040 Wien, 
Austria}\thanks{Supported by FWF under P13997-TPH.}
and
H. Reinhardt\addressmark[TUB]\thanks{Supported by DFG under
Re 856/4-1.}
}

\begin{document}

\begin{abstract}
A model for the infrared sector of $SU(2)$ Yang-Mills
theory, based on magnetic vortices represented by (closed) random surfaces,
is presented. The model quantitatively describes both confinement
(including the finite-temperature transition to a deconfined phase) and
the topological susceptibility of the Yang-Mills ensemble. A first
(quenched) study of the spectrum of the Dirac operator furthermore yields a
behavior for the chiral condensate which is compatible with results obtained 
in lattice gauge theory.
\vspace{0.2cm}
\end{abstract}

\maketitle

Diverse nonperturbative effects characterize strong interaction physics.
Color charge is confined, chiral symmetry is spontaneously broken, and
the axial $U(1)$ part of the flavor symmetry exhibits an anomaly.
Various model explanations for these phenomena have been advanced;
to name but two widely accepted ones, the dual superconductor 
mechanism of confinement, and instanton models, which describe the 
$U_A (1)$ anomaly and spontaneous chiral symmetry breaking. However, no 
clear picture has emerged which comprehensively describes
infrared strong interaction physics within one common framework. The vortex 
model presented here~\cite{selprep,preptop} aims to bridge this gap. 
On the basis of a simple effective dynamics, it simultaneously reproduces 
the confinement properties of $SU(2)$ Yang-Mills theory (including 
the finite-temperature deconfinement transition), as well as the 
topological susceptibility, which encodes the $U_A (1)$ anomaly.
Furthermore, a first (quenched) study of the spectrum of the Dirac 
operator yields a behavior for the chiral condensate which is compatible 
with results obtained in lattice gauge theory; this indicates that the 
model also correctly reproduces the spontaneous breaking of chiral symmetry.

Center vortices are closed chromomagnetic flux lines in three-dimensional
space; thus, they are described by closed two-dimensional world-surfaces
in four-dimensional space-time. In the $SU(2)$ case, their magnetic flux is 
quantized such that they modify any Wilson loop by a phase factor $(-1)$
when they pierce an area spanned by the loop.
To arrive at a tractable vortex model, it is useful to compose the vortex 
world-surfaces out of plaquettes on a hypercubic lattice. The spacing of 
this lattice is a fixed physical quantity (related to a thickness of the 
vortex fluxes), and represents the ultraviolet cutoff inherent
in any infrared effective framework. The model vortex surfaces are
regarded as random surfaces, and an ensemble of them is generated using
Monte Carlo methods. The corresponding weight function penalizes curvature
by associating an action increment $c$ with every instance of two
plaquettes which are part of a vortex surface, but which do not lie 
in the same plane, sharing a link (note that several such pairs of
plaquettes can occur for any given link).

\begin{figure}[h]
\centerline{
\epsfxsize=7.5cm
\epsffile{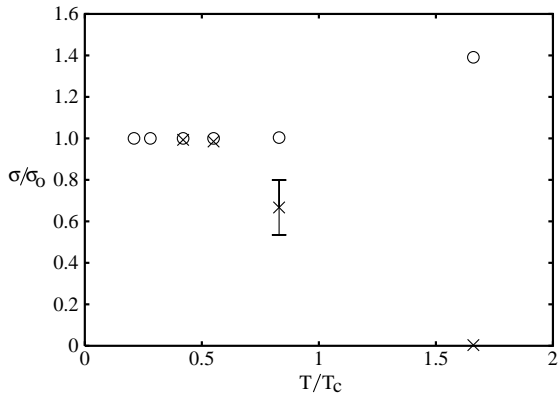} 
\vspace{-1cm}
}
\caption{Confinement properties as a function of temperature, with $c=0.24$.
Crosses: string tension between static quarks; circles: spatial 
string tension $\sigma_{s} $. Whereas the former
has largely been fitted using the freedom in the choice of $c$ (see text), 
the latter is predicted. In the deconfined regime, it begins to rise with 
temperature; the value $\sigma_{s} (T=1.67 \, T_C ) = 1.39 \, \sigma_{0} $ 
reproduces the one obtained in full $SU(2)$ Yang-Mills
theory {\protect \cite{karsch}}.}
\label{stt}
\end{figure}

Via the definition given above, Wilson loops (and, in complete analogy,
Polyakov loop correlators) can be evaluated in the vortex ensemble, and 
string tensions extracted. For sufficiently small curvature coefficient 
$c$, one finds a confined phase (non-zero string tension) at low 
temperatures, and a transition to a high-temperature deconfined phase. 
For $c=0.24$, the $SU(2)$ Yang-Mills relation between the deconfinement
temperature and the zero-temperature string tension,
$T_C /\sqrt{\sigma_{0} } =0.69$, is reproduced. When furthermore setting 
$\sigma_{0} =(440 \, \mbox{MeV} )^2 $ to fix the scale,
measurement of $\sigma_{0} a^2 $ yields the lattice spacing 
$a=0.39 \, \mbox{fm} $. The full temperature dependence of the
string tensions is displayed in Fig.~\ref{stt}.
\vspace{-0.7cm}

\vspace{-0.3cm}
Note that the confined and deconfined phases can alternatively be 
characterized by certain percolation properties \cite{selprep,tlang} 
of the vortex clusters.

\begin{figure}[h]
\centerline{
\epsfxsize=7.5cm
\epsffile{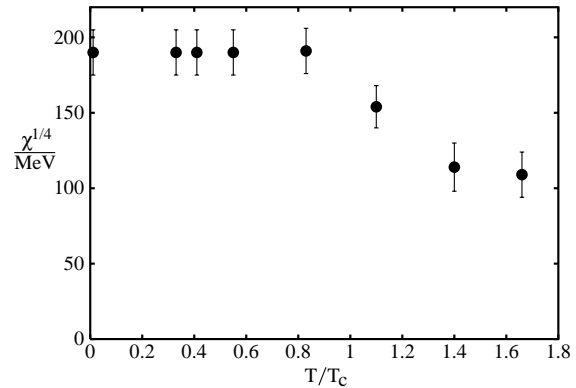}
\vspace{-1cm}
}
\caption{(Fourth root of) the topological susceptibility
as a function of temperature, with $c=0.24$.
Also this result is quantitatively compatible
with the one in full Yang-Mills theory {\protect \cite{digia}}.}
\label{tss}
\end{figure}

Complementarily, also the topological properties of the Yang-Mills ensemble 
encode important nonperturbative effects. The topological charge $Q$ of a 
vortex surface configuration is carried by its singular 
points \cite{cont,preptop,protop}, i.e.~points at which the set of tangent 
vectors to the surface configuration spans all four space-time directions 
(a simple example are surface self-intersection points). Since a vortex 
surface carries a field strength characterized by a nonvanishing tensor 
component associated with the two space-time directions locally orthogonal 
to the surface \cite{cont}, these singular points are precisely the points 
at which the topological charge density 
$\epsilon_{\mu \nu \lambda \tau } \, \mbox{Tr} \, F_{\mu \nu }
F_{\lambda \tau } $ is non-vanishing. Note that, in order to carry
nontrivial global topological charge, surfaces must be non-oriented;
lines along which the orientation flips can be associated with Abelian
magnetic monopoles \cite{cont}. Generic vortex surfaces in the ensemble
studied here are indeed non-orientable. In practice, identifying all
singular points of the hypercubic lattice surfaces used in the present 
model involves resolving ambiguities~\cite{preptop} reminiscent of those
contained in lattice Yang-Mills link configurations. The resulting
topological susceptibility $\chi = \langle Q^2 \rangle /V$, where $V$
denotes the space-time volume under consideration, is exhibited in 
Fig.~\ref{tss}.
\vspace{-0.7cm}

\begin{figure}[h]
\centerline{
\epsfxsize=7.5cm
\epsffile{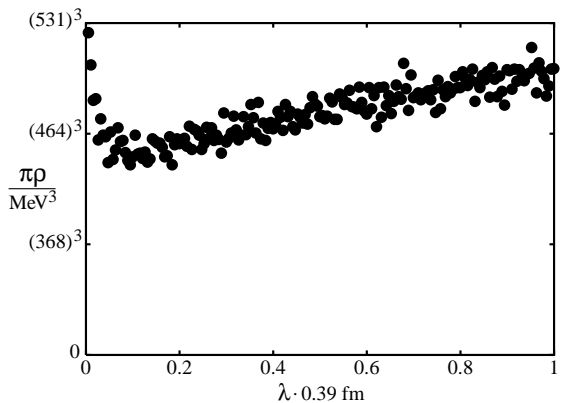}
\vspace{-1cm}
}
\caption{Dirac spectral density $\rho (\lambda )$ in a universe of volume 
$(1.17 \, \mbox{fm} )^4 $, with $c=0.24$.}
\label{dspe}
\end{figure}

\vspace{-0.3cm}
A comprehensive description of the nonperturbative phenomena which
determine strong interaction physics must furthermore include the
coupling of the vortices to quark degrees of freedom and the associated 
spontaneous breaking of chiral symmetry. The latter can be quantified via 
the chiral condensate $\langle \bar{\psi } \psi \rangle $, which is
related to the spectral density $\rho (\lambda )$ of the Dirac operator
in a vortex background via the Casher-Banks formula
$\langle \bar{\psi } \psi \rangle = \pi \rho (0)$.
One can locally associate the chromomagnetic flux represented by an 
arbitrary vortex surface with a continuum gauge field \cite{cont}; this 
allows to construct the Dirac operator directly in the continuum, 
preserving exact chiral symmetry. Globally, one must allow for the gauge 
field to be defined on different space-time patches, in order to
accomodate the non-orientability of the vortex surfaces. While the gauge 
fields on each patch can be chosen Abelian (i.e.~in 3-direction in color 
space), the transition functions in the overlap regions in general have 
to be non-Abelian. Since the spectrum of the Dirac operator here is
obtained using the finite element method, it is natural to use the 
different finite elements as the space-time patches. A sample Dirac
spectrum generated using this method is displayed in Fig.~\ref{dspe}.
\vspace{-0.7cm}

\vspace{-0.3cm}
Note the anomalous enhancement of $\rho (\lambda )$ near $\lambda =0$, in
accordance with the divergence
expected for a Dirac operator with good chiral properties \cite{chen}.
Also the linear extrapolation of the bulk of the spectrum 
to $\lambda =0$, i.e.~the spectral density obtained 
by truncating the $\lambda =0$ divergence, quantitatively lies in the 
range obtained in full $SU(2)$ Yang-Mills theory \cite{hatep}.
Of course, the chiral condensate by itself does not directly represent a
renormalization group invariant, physical quantity, but can only
interpreted in conjunction with a definition of the current quark
masses. These can only be reliably fixed in the present model after
more detailed (ultimately unquenched) measurements of hadronic properties.

It is also instructive to consider a change in the construction 
of the gauge field which trivializes the topology. This can be
done by replacing the monopole loops on the vortex surfaces 
by double vortex sheets spanning areas bounded by the loops. Then,
the surfaces can be oriented, the global topological charge vanishes,
and presumably also many local clusters of nontrivial topology are
trivialized (although the topological density as such does not
necessarily vanish). In the Dirac spectrum, this leads to a truncation
of the $\lambda =0$ divergence, and instead the finite-volume gap
near $\lambda =0$ familiar e.g.~from chiral random matrix theory opens,
without the bulk of the spectrum being qualitatively changed.
While the $\lambda =0$ divergence thus seems
intimately related to topological properties \cite{shatep}, chiral
symmetry breaking as such merely appears to require randomness of the gauge
fields in a very general sense, without specific reference to
topology.

Several improvements of the present model are envisaged. For one, the
treatment must be generalized from $SU(2)$ to $SU(3)$ color. Also the
Dirac operator can still be developed further to explicitly take into 
account a finite thickness of the vortices; the above
construction, while embodying the correct
topology, still deals with magnetic flux localized on infinitely
thin world-surfaces. Eventually, it is hoped that this model will
become a useful tool for phenomenological considerations.

\end{document}